\documentclass[aps,preprint,pre,floatfix,nofootinbib,superscriptaddress,12pt]{revtex4-1}
\usepackage{graphicx}
\usepackage{bm}
\usepackage{amssymb}
\usepackage{amsmath}
\usepackage{textcomp}
\usepackage{bibentry}
\usepackage[dvips,usenames]{color}
\usepackage{enumerate}

\begin{document}

\title{
Correction of Doi-Edwards' Green Function for a Chain in a Harmonic Potential
and its Implication for the Stress-Optical Rule
}

\author{Jay D. Schieber}
\affiliation{\small{
Department of Chemical and Biological Engineering,\\
Department of Physics, and \\
Center for Molecular Study of Condensed Soft Matter \\
Illinois Institute of Technology \\
3440 S. Dearborn St., Chicago, Illinois 60616, U.S.A. \\
}}

\author{Tsutomu Indei}
\affiliation{\small{
Department of Chemical and Biological Engineering, and \\
Center for Molecular Study of Condensed Soft Matter \\
Illinois Institute of Technology \\
3440 S. Dearborn St., Chicago, Illinois 60616, U.S.A. \\
}}

\begin{abstract}

We provide a corrected Green's function for a polymer chain trapped in 
a two-dimensional anisotropic harmonic potential with a fixed boundary condition.
This Green's function is a modified version of what Doi and Edwards first derived to describe the polymer chain 
confined in the tube-like domain of surrounding entangled polymers [J. Chem. Soc. Farad. Trans. II 74 (1978) 1802].
In contradiction to the results found by Ianniruberto and Marrucci (IM)
\textcolor{black}{when applying the Doi-Edwards Green function} 
[J. Non-Newtonian Fluid Mech. 79 (1998) 225],
we find that the stress-optical rule is violated for any tube potential either circular or elliptic
if the corrected Green's function is used.
The violation is due to the presence of the virtual springs
to confine the chain in the tube rather than the anisotropy of the confinement potential.  
On the other hand, Doi and Edwards used their Green's function only for estimation of the monomer density along the primitive path 
where we find just a small correction. 
Since they did not use it for rheological calculations, the stress-optic rule appears to be safe for the Doi-Edwards model. 

\end{abstract}

\maketitle

\section{Introduction}

For polymer melts, it is well established experimentally that 
there is a proportionality relationship between the stress tensor and the refractive index tensor 
as long as the polymer chains are Gaussian 
\cite{Kriegl_PMRFB,Fuller_ORCF,Venerus_JOR_1999,Luapa_JNNFM_2006}.
This linear relation is called the stress-optical rule (SOR) 
\cite{Fuller_ORCF}.
For Gaussian chains, the traction force of a chain strand 
whose ends are constrained by entanglements 
is a linear function of the strand's end-to-end vector, so that
the stress tensor of the polymer melt is proportional to the 
second-moment of the strand's orientation vector \cite{Bird_DPL}.
On the other hand, the refractive index tensor is proportional to the same second moment tensor,
so that the SOR holds for Gaussian chains.
If the polymer chain is stretched beyond the Gaussian regime, 
SOR is violated because of the breakdown of the linear relation between the traction and the end-to-end length of the strand
\cite{Venerus_JOR_1999,Luapa_JNNFM_2006}.

The tube model of entangled polymers assumes that the lateral motion of the polymer chain in the melt is prohibited, 
as if the chains were confined in a tube-like domain \cite{Doi_TPD}.
The origin of the stress of the material is usually assumed to be
the traction force of the chain strand along the tube.
But in 1998, Ianniruberto and Marrucci (IM) considered the possibility
that the pressure exerted by the chain strand on the confining tube wall
may contribute significantly to the stress tensor
together with the traction along the tube \cite{IM_JNNFM_1998}.
They then investigated theoretically whether or not the SOR is obeyed in such a case.
Based on the assumption that the equilibrium axial symmetry of the tube will be destroyed by the deformation of the melt,
and the local geometry of the tube segment will become biaxial,
IM considered the situation that the tube cross section is elliptical, or, in other words, 
the lateral pressure on the tube wall is anisotropic.
Effects of the lateral pressure were taken into account by introducing
``virtual springs" that connect all parts of the chain strand to the central axis of the tube.
Then IM calculated the stress tensor and the polarizability tensor by applying a slightly modified
Green's function that was first derived by Doi and Edwards (DE) 
to estimate the monomer density per length of the tube with circular cross-section \cite[App.~A]{Doi_Edwards_78b}.
With the help of a ``free confinement assumption" (more about this below) that each 
link of the chain does not suffer from the confinement potential,
IM purportedly proved that the SOR is obeyed even when the contribution from the anisotropic lateral pressure 
(or confinement potential of the virtual springs) contributes to the stress.

However, the Green functions given by DE and applied by IM are actually incorrect because these do not satisfy the 
boundary condition \textcolor{black}{or continuity condition} (see Sec.~\ref{sec:incorrectGreen}).
These Green functions can approximately describe an extreme case where the confinement potential is so large
that the contribution of the chain traction along the tube cross-section is negligible
compared to that of the confinement potential (see Sec.~\ref{sec:correctGreen}).
But IM did not take this limitation into consideration in their attempt to prove the SOR \cite{IM_JNNFM_1998}.

In this paper, we derive the corrected Green function 
so that the boundary \textcolor{black}{(continuity)} condition is satisfied (in Sec.~\ref{sec:correctGreen}), 
and reconsider if the SOR is obeyed in IM's model of entangled polymer melts.
We show in Sec.~\ref{sec:violateSOR} that, if the polarizability tensor and the stress tensor
are calculated based on the corrected Green function, and also
if the calculation is appropriately conducted without employing the free confinement assumption, then
the SOR is not obeyed due to the presence of the confinement potential
--- regardless of whether the confinement potential is isotropic or anisotropic 
(i.e., the cross-section of the tube is circular or ellipsoidal)
and also irrespective of whether the confinement is strong or weak.
The deviation of the stress tensor from the polarizability tensor is
exactly the stress components originating from the virtual springs of the confinement potential.

DE estimated the monomer density along the primitive path \cite[App.~A]{Doi_Edwards_78b} based on their Green function.
Although it does not satisfy the boundary \textcolor{black}{(continuity)} condition,
the DE prediction of the monomer density is valid as shown in Sec.\ref{sebsec:monomerdensity}.
On the other hand, DE did not use the Green function for rheological calculations,
so that the SOR appears to be safe in the DE model for rheology predictions.

Use of the corrected Green function might also be of considerable importance in microscopic studies of the confining tube potential.  There have been recent attempts to model the confining potential numerically \cite{KrogerRamirezEtAl_2002}, find the potential numerically through molecular dynamics \cite{ZhouLarson_2006}, or predict the potential from a more microscopic basis \cite{ReadEtAl_2008,SussmanSchweizer_2012,SteenbakkersSchieber_2012}. Some of these works have suggested that the potential might be anharmonic, in contradiction with what was assumed by Doi and Edwards and here.  There is also interest in how the potential might be affected by flow \cite{Schieber_et_al_03,BaigMavrantzasEtAl_2010}.  Therefore, it should be important to have the correct statistics arising from the simple harmonic assumption first to facilitate mapping from the atomistic level.

In this paper, we use the same notation as in IM's paper \cite{IM_JNNFM_1998} except that the components
of the vector $\bm{R}$ are denoted as $(R_x,R_y,R_z)$ instead of $(x,y,z)$, 
and the unit vectors along each axis are represented by ($\bm{\delta}_x,\bm{\delta}_y,\bm{\delta}_z$)
instead of $(\hat{\bm{x}},\hat{\bm{y}},\hat{\bm{z}})$.

\section{Doi-Edwards' Green function for chain confinement in a harmonic potential}
\label{sec:incorrectGreen}

For the purpose of estimating the monomer density of an entangled polymer strand per tube length,
DE described the confinement of the chain in the tube
by the two-dimensional isotropic harmonic potential given by \cite[App.~A]{Doi_Edwards_78b}
\begin{eqnarray}
V=\frac{k_{\rm B}T}{6}w^2(R_x^2+R_y^2)
\label{eq:potential}
\end{eqnarray}
where $k_{\rm B}T$ is the thermal energy, $w$ represents the confinement strength, and $\bm{R}$ is the position of a point 
(called the bead or monomer according to IM, though the chain is described as a continuum object) on the chain. 
Then the Green function $G(\bm{R},\bm{R}_0;n,n_0)$ satisfying
\begin{eqnarray}
\left(\frac{\partial}{\partial n}-\frac{b^2}{6}\frac{\partial^2}{\partial \bm{R}^2}+\frac{V}{k_{\rm B}T}\right)
G(\bm{R},\bm{R}_0;n,n_0)=\delta(n-n_0)\delta(\bm{R}-\bm{R}_0)
\label{eq:Green}
\end{eqnarray}
gives the statistical probability that the $n$th Kuhn step is near $\bm{R}$, for a strand trapped in the tube by the potential (\ref{eq:potential}), given that the $n_0$th bead is at $\bm{R}_0$.  
The delta functions in the right-hand side take account of the boundary conditions
$G(\bm{R},\bm{R}_0;n,n_0)=0$ (for $n\neq n_0$) and $G(\bm{R},\bm{R}_0;n_0,n_0)=\delta(\bm{R}-\bm{R}_0)$
\cite[p.17-19]{Doi_TPD}.
It is important to note that the second delta function on the right-hand side of Eq.~(\ref{eq:Green}) makes the chain continuous 
\cite[Sec.3]{Gardiner}.
$b$ is the persistence length of the chain, and also represents the strength of the connector springs.
We call the portion of the chain with the persistent length $b$ the link, as did IM.
DE used the Green function in Ref.~\cite[App.A]{Doi_Edwards_78b} to show that the equilibrium monomer density 
per tube length is of the order of $a/b^2$, where $a$ is the dimension of the tube cross-section.
We note that the conformation of the confined strand
is governed by the Ornstein-Uhlenbeck process \cite{Uhlenbeck_PR_1930}
when $n$ is interpreted as time because the confinement potential is harmonic.

DE gave the following expression \cite[Eq.~(A.6)]{Doi_Edwards_78b} as a solution to Eq.~(\ref{eq:Green})
\footnote{A factor 2 in the denominator of the first term of the exponential function
is missing in Eq.~(A.6) of Ref.~\cite{Doi_Edwards_78b},
and a prefactor of the exponential function in Eq.~(2) of Ref.~\cite{IM_JNNFM_1998} is a typo.
Also the misprint correction seems not to be reported in reference [14] of Ref.~\cite{IM_JNNFM_1998}.
Note that the solution is used for arbitrary $\bm{R},\bm{R}',n,n'$, not just for $\bm{R}'=\bm{0},n'=0$.}:
\begin{eqnarray}
G^{\rm (DE)}(\bm{R},\bm{0};n,0)
\propto \frac{1}{n^{1/2}}\exp\left[-\frac{w}{2b}(R_x^2+R_y^2)-\frac{3R_z^2}{2nb^2}-\frac{nbw}{3}\right].
\label{eq:GsolutionDE}
\end{eqnarray}
But Eq.~(\ref{eq:GsolutionDE}) does not satisfy the boundary condition 
$G(\bm{R},\bm{0};0,0)=\delta(\bm{R})$ for the $x$ and $y$ components of $\bm{R}$,
thereby violating the continuity of the chain.
Moreover, in the absence of the confinement potential ($w=0$), 
Eq.~(\ref{eq:GsolutionDE}) becomes uniform for $x$ and $y$ components 
while both ends are pinned (which is only possible because of the discontinuity of the chain).
That is, while the $z$-coordinates of the monomer are continuous, the other coordinates are not, so that
monomers are free to fly apart along the direction perpendicular the tube central axis.
(If $w$ is large, they can stay relatively close, but the chain is still discontinuous.)
\textcolor{black}{These are results of the fact}
that the effects of traction along the $x$ and $y$ axes
are not included in Eq.~(\ref{eq:GsolutionDE});
only influences of the confinement 
are taken into account.
Thus Eq.~(\ref{eq:GsolutionDE}) leads to the wrong stress and polarizability tensors
that do not include the effects of traction perpendicular to the tube central axis.
The derived two tensors are appropriate only when the confinement is much stronger than the traction.
This limitation should be kept in mind whenever one uses or applies Eq.~(\ref{eq:GsolutionDE}).

\section{Corrected Green function for anisotropic harmonic potential}
\label{sec:correctGreen}

Here we find the Green function that satisfies Eq.~(\ref{eq:Green}).
Like IM, we consider the general, anisotropic chain confinement whose potential is described by
\begin{eqnarray}
V=\frac{k_{\rm B}T}{6}(w_x^2R_x^2+w_y^2R_y^2).
\label{eq:potentiala}
\end{eqnarray}
Since the $x,y,z$-components are decoupled,
the solution of Eq.~(\ref{eq:Green}) can be decomposed as
\begin{eqnarray}
G(\bm{R},\bm{R}_0;n,n_0)=\prod_{\beta=x,y,z}G_{\beta}(R_{\beta},R_{0,\beta};n,n_0).
\label{eq:G3D}
\end{eqnarray}
We show that the solutions for the perpendicular components $\beta=x,y$ is 
(see the following subsection \ref{sec:derivation})
\begin{eqnarray}
G_{\beta}(R_{\beta},R_{0,\beta};n,n_0)=J_{\beta}
\exp \left(-\frac{w_{\beta}}{2b}
\coth \left[\frac{(n-n_0)bw_{\beta}}{3}\right] 
\left(R_{\beta}-\overline{R}_{\beta}\right)^2\right)
~~~~~(\mbox{for}~\beta=x,y),~~~
\label{eq:Gsolution}
\end{eqnarray}
where the prefactor is given by
\begin{eqnarray}
J_{\beta}=\sqrt{\frac{w_{\beta}}{2\pi b}
\text{csch}\left[\frac{(n-n_0)bw_{\beta}}{3}\right]}
\exp\left(-\frac{w_{\beta}}{2 b} 
\coth \left[\frac{(n-n_0)bw_{\beta}}{3}\right]
\left(R_{0,\beta}^2-\overline{R}_{\beta}^2\right)\right),
\end{eqnarray}
and the first moment is
\begin{eqnarray}
\overline{R}_{\beta}
=R_{0,\beta} \text{sech}\left[\frac{(n-n_0)bw_{\beta}}{3}\right].
\label{eq:firstmoment}
\end{eqnarray}
For the parallel component $\beta=z$, the solution is
\begin{eqnarray}
G_z(R,R_{0,z};n,n_0)=
\sqrt{\frac{3}{2\pi (n-n_0)b^2}}
\exp\left(-\frac{3\left(R_z-R_{0,z}\right)^2}
{2(n-n_0)b^2}\right).
\label{eq:Gzsolution}
\end{eqnarray}

One can see that Eq.~(\ref{eq:Gsolution}) is a plausible solution of Eq.~(\ref{eq:Green})
for the following reasons.
Firstly, in the weak confinement limit $w_\beta\to 0$ (for $\beta=x,y$), Eq.~(\ref{eq:Gsolution}) becomes
the same form as Eq.~(\ref{eq:Gzsolution}), i.e.,
\begin{eqnarray}
G_{\beta}(R,R_{0,\beta};n,n_0)=
\sqrt{\frac{3}{2\pi |n-n_0|b^2}}
\exp\left(-\frac{3\left(R_{\beta}-R_{0,\beta}\right)^2}
{2|n-n_0|b^2}\right),
\label{eq:Gsolutionlimit}
\end{eqnarray}
as expected for a Rouse chain \cite[\S2.3]{Doi_Edwards_86}.  Secondly, if $n$ is close to $n_0$ for a fixed finite $w_\beta$,
Eq.~(\ref{eq:Gsolution}) is also approximately written as Eq.~(\ref{eq:Gsolutionlimit}).
Thus, in the limit of $n\to n_0$, Eqs.~(\ref{eq:Gsolution}) as well as (\ref{eq:Gzsolution}) become 
$\delta(R_\beta-R_{0,\beta})$~(for $\beta=x,y,z$),
thereby satisfying the boundary condition
$G_{\beta}(R,R_{0,\beta};n_0,n_0)=\delta(R-R_{0,\beta})$ for all components.
And thirdly, if the inequality 
\begin{eqnarray}
nbw_\beta\gg1~~~~~~(\mbox{for}~\beta=x,y)
\label{eq:inequ}
\end{eqnarray}
is satisfied, then Eq.~(\ref{eq:G3D}) or
\begin{eqnarray}
G(\bm{R},\bm{0};n,0)=
\sqrt{\frac{3}{2\pi n b^2}
\prod_{\beta=x,y}\frac{w_{\beta}}{2\pi b}
\text{csch}\left[\frac{nbw_{\beta}}{3}\right]}
\exp \left[-\sum_{\beta=x,y}\frac{w_{\beta}}{2b}
\coth \left[\frac{nbw_{\beta}}{3}\right]R_{\beta}^2-\frac{3R_z^2}{2nb^2}\right],~~~~
\label{eq:Gsolution0}
\end{eqnarray}
becomes 
\begin{eqnarray}
G^{\rm {(IM)}}(\bm{R},\bm{0};n,0)
\propto\frac{1}{n^{1/2}}
\exp\left[-\frac{w_x}{2b}R_x^2-\frac{w_y}{2b}R_y^2-\frac{3R_z^2}{2nb^2}-\frac{nb(w_x+w_y)}{6}\right]
~~~(nbw_\beta\gg1)~~~~
\label{eq:GIM}
\end{eqnarray}
where, for simplicity, we put $n_0=0$ and $\bm{R}_0=0$ without loss of generality.
Equation (\ref{eq:GIM}) is the Green function that IM used (Eq.~(2) of Ref.~\cite{IM_JNNFM_1998})
to derive the polarizability and stress tensors.
For an isotropic potential ($w_x=w_y$), Eq.~(\ref{eq:GIM}) reduces to Eq.~(\ref{eq:GsolutionDE}).
Thus the Green function of IM (or DE for the isotropic case) is appropriate only when condition (\ref{eq:inequ}) is satisfied.

The inequality (\ref{eq:inequ}) has two interpretations.
For a given strength $bw_\beta$ of the confinement potential, 
Eq.~(\ref{eq:inequ}) indicates that $n$ 
is so large as to reach the asymptotic state
(or the `steady state' of the Ornstein-Uhlenbeck process if $n$ is interpreted as time) 
where the effect of the boundary condition is negligible.
On the other hand, if $n$ is given, Eq.~(\ref{eq:inequ}) implies that the strength of the
confinement potential is so strong as to overcome the traction for the $x$ and $y$ components. 
DE apparently neglected the traction along the $x$ and $y$ axes to obtain Eq.~(\ref{eq:GsolutionDE}).

\subsection{Derivation of Eq.~(\ref{eq:Gsolution})}
\label{sec:derivation}

$G_{\beta}$ is expected to be Gaussian because the potential is harmonic.
Therefore we can assume that it has the expression
\begin{eqnarray}
G_{\beta}=\exp\Bigl(-f(n)R_{\beta}^2-g(n)R_{\beta}-h(n)\Bigr).
\label{eq:Gexp}
\end{eqnarray}
\textcolor{black}{Now we substitute} this expression into the $\beta(=x,y)$-component of Eq.~(\ref{eq:Green}) 
but without the delta functions (which are enforced below)
\begin{eqnarray}
\left(\frac{\partial}{\partial n}-\frac{b^2}{6}\frac{\partial^2}{\partial R_\beta^2}
+\frac{w_\beta^2}{6}R_{\beta}^2\right)
G_{\beta}=0.
\label{eq:Green2}
\end{eqnarray}
Equating like powers of $R_\beta$, we obtain three ordinary differential equations
\begin{subequations}
\begin{eqnarray}
& &\frac{df(n)}{dn}+\frac{2b^2}{3}f(n)^2-\frac{w_{\beta}^2}{6}=0, \label{eq:fn}\\
& &\frac{dg(n)}{dn}+\frac{2b^2}{3}f(n)g(n)=0, \label{eq:gn}\\
& &\frac{dh(n)}{dn}-\frac{b^2}{3}f(n)+\frac{b^2}{6}g(n)^2=0. \label{eq:hn}
\end{eqnarray}
\end{subequations}
Equation~(\ref{eq:fn}) is the Riccati equation for $f(n)$ and has solution
\begin{eqnarray}
f(n)=\frac{w_\beta}{2b}\text{coth}\left[\frac{(n-n_0)bw_\beta}{3}\right]
\label{eq:fsolution}
\end{eqnarray}
where $n_0$ is the constant.
It approaches $f(n)\to\frac{3}{2(n-n_0)b^2}$ in the unconstrained limit $w_{\beta}\to0$, which is appropriate.
We note that there is another solution of Eq.~(\ref{eq:fn}) 
\begin{eqnarray}
f(n)=\frac{w_\beta}{2b}\text{tanh}\left[\frac{(n-n_0)bw_\beta}{3}\right].
\end{eqnarray}
However, this solution goes to 0 at $w_{\beta}\to0$,
and is therefore unphysical
since the delta-function boundary conditions are not satisfied.
Therefore we employ Eq.~(\ref{eq:fsolution}) as the solution of Eq.~(\ref{eq:fn}).
Then $g(n)$ and $h(n)$ are obtained from Eqs.~(\ref{eq:gn}) and (\ref{eq:hn}) as
\begin{eqnarray}
g(n)=C\frac{w_\beta}{2b}\text{csch}\left[\frac{(n-n_0)bw_\beta}{3}\right]
\end{eqnarray}
and
\begin{eqnarray}
h(n)=
\frac{1}{2} \log \left(\sinh\left[\frac{(n-n_0)bw_{\beta}}{3}\right]\right)
+C^2\frac{b}{2 w_{\beta}} \coth \left[\frac{(n-n_0)bw_{\beta}}{3}\right]
\end{eqnarray}
respectively, where $C$ is a constant.
Now that we have $G_{\beta}$,
the first moment of $R_{\beta}$ is calculated as
\begin{eqnarray}
\overline{R}_{\beta}
:=\frac{\int_{-\infty}^{\infty}RG_{\beta}(R)dR}
{\int_{-\infty}^{\infty}G_{\beta}(R)dR}
=-C\frac{b}{w_{\beta}}\text{sech}\left[\frac{(n-n_0)bw_{\beta}}{3}\right].
\end{eqnarray}
Thus, by putting $n=n_0$, the constant $C$ is determined as
\begin{eqnarray}
C=-R_{0,\beta}\frac{w_{\beta}}{b},
\end{eqnarray}
and then the first moment is given by Eq.~(\ref{eq:firstmoment}).
Multiplying Eq.~(\ref{eq:Gexp}) by the constant $\sqrt{\frac{w_{\beta}}{2\pi b}}$ 
so that the boundary condition is satisfied, $G_{\beta}$ is written as Eq.~(\ref{eq:Gsolution}).
The $z$-component (Eq.~(\ref{eq:Gzsolution})) is obtained by replacing $\beta$ with $z$ in Eq.~(\ref{eq:Gsolution})
and by taking the limit $w_z\to0$.

\section{Application of the correct Green function}
\label{sec:violateSOR}

In this section, we calculate several quantities by applying the Green function given by Eq.~(\ref{eq:Gsolution}),
and examine the SOR.

\subsection{Number of monomers in the tube}

\begin{figure}[t]
\begin{center}
\includegraphics*[scale=0.5]{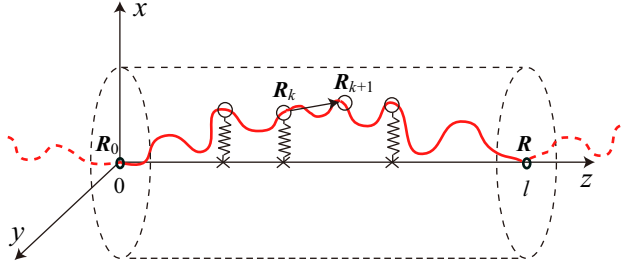}
\end{center}
\vspace*{-0.5cm}
\caption{
A chain strand confined in the tube-like region.
Confinement potential due to the tube, described by the virtual springs in this figure,
affects all portion of the segment in a continuous manner.
Ends of the chain strand are pinned on the central $z$-axis of the tube segment.
}
\label{fig:tube}
\end{figure}

\begin{figure}[t]
\begin{center}
\includegraphics*[scale=0.4]{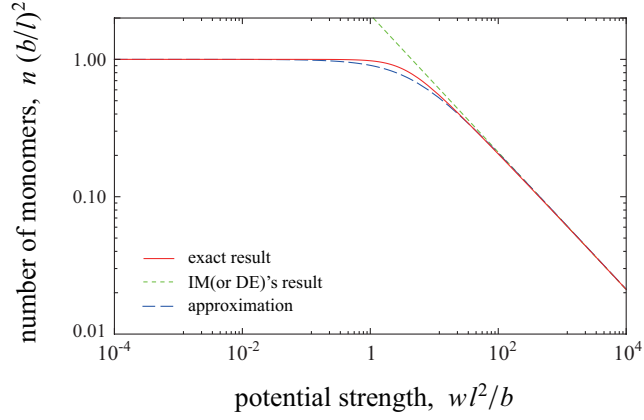}
\end{center}
\vspace*{-0.5cm}
\caption{
The equilibrium number of monomers of a chain strand in the tube 
plotted against the strength of the confinement potential.
For simplicity an isotropic potential $w_x=w_y(=:w)$ is considered.
Solid line is the \textcolor{black}{exact result (numerical solution of Eq.~(\ref{eq:monoden}))}, 
dotted line is the IM (or DE) result (Eq.~(\ref{eq:monoden1})), 
and dashed line is the approximate result (Eq.~(\ref{eq:monodenap})).
}
\label{figmono}
\end{figure}

We consider the simple case that the start and end of the chain strand
are pinned at the central axis of the tube separated by the distance $l$, as 
both DE and IM assumed (see Fig.~\ref{fig:tube}).
By putting $\bm{R}_0=\bm{0}$ and $\bm{R}=l\bm{\delta}_z$ in Eq.~(\ref{eq:Gsolution}) 
(or substituting $R_x=R_y=0,~R_z=l$ in Eq.~(\ref{eq:Gsolution0})),
we have
\begin{eqnarray}
G(\bm{R}{=l\bm{\delta}_z},\bm{0},n,0)=
\exp\left[
\frac{1}{2}\sum_{\beta=x,y}
\log\left(
\frac{w_{\beta}}{2\pi b}
\text{csch}\left[\frac{nbw_{\beta}}{3}\right]
\right)
-\frac{3l^2}{2nb^2}
+\frac{1}{2}\log\left(\frac{3}{2\pi n b^2}\right)
\right].~~~
\label{eq:simpleG}
\end{eqnarray}
The most probable value of $n$ is obtained by maximizing Eq.~(\ref{eq:simpleG}) with respect to $n$ for a fixed $l$ 
\cite{Doi_Edwards_78b}.
If we take the derivative of the argument of the exponential function with respect to $n$
and set the result to 0, we obtain
\begin{eqnarray}
n=\frac{3l}{b\sqrt{b\sum_{\beta=x,y}w_{\beta}\coth\left[\frac{nbw_{\beta}}{3}\right]
+\frac{3}{n}}}.
\label{eq:monoden}
\end{eqnarray}
In the condition that the inequality (\ref{eq:inequ}) is satisfied,
$\coth[nbw_\beta/3]$ is approximately 1, and $3/n$ in the denominator can be neglected.
Then the most probable value of $n$ that
DE \cite[Eq.~(A.9)]{Doi_Edwards_78b} and IM \cite[Eq.~(3)]{IM_JNNFM_1998} derived 
for the isotropic and anisotropic potentials, respectively, is obtained, i.e.,
\begin{eqnarray}
n\simeq \frac{3l}{b\sqrt{2b\bar{w}}}~~~~~~~~~~~(\mbox{for}~nbw_\beta\gg1)
\label{eq:monoden1}
\end{eqnarray}
where we put $\bar{w}:=(w_x+w_y)/2$ as the mean strength of the confinement potential.
On the other hand, in the weak confinement limit, Eq.~(\ref{eq:monoden}) 
becomes
\begin{eqnarray}
n\to\frac{l^2}{b^2}~~~~~(w_\beta\to0).
\label{eq:monoden0}
\end{eqnarray}
An expression that interpolates Eqs.~(\ref{eq:monoden1}) and (\ref{eq:monoden0}) gives an approximate
solution to Eq.~(\ref{eq:monoden}) as
\begin{eqnarray}
n\simeq \frac{3l}{b\sqrt{2b\bar{w}+9b^2/l^2}}.
\label{eq:monodenap}
\end{eqnarray}
See Fig.~\ref{figmono} for reference.
By comparing Eqs.~(\ref{eq:monoden1}) and (\ref{eq:monoden0}),
the crossover of $w$ between the weak and strong confinement can be estimated as 
$w_c\simeq \frac{9b}{2l^2}$.
If the condition $w_\beta\ll w_c$ is satisfied, then the inequality opposite of Eq.~(\ref{eq:inequ}) is fulfilled 
because $nbw_\beta \ll nb^2/l^2\simeq 1$.

\subsection{Tube dimension}

\begin{figure}[t]
\begin{center}
\includegraphics*[scale=0.4]{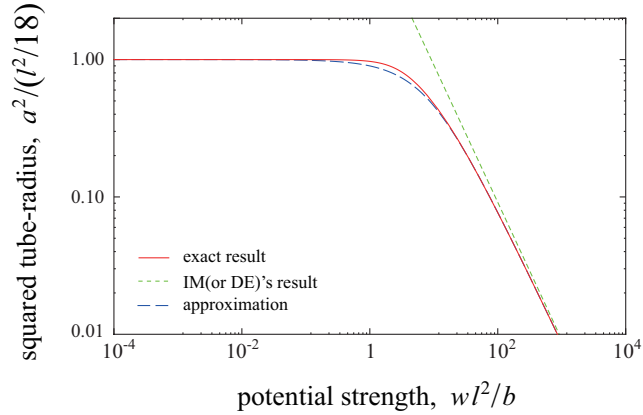}
\end{center}
\vspace*{-0.5cm}
\caption{
Squared radius of gyration of the tube cross section along the axis
perpendicular to the tube plotted against the strength of the confinement potential.
For simplicity an isotropic potential is considered $(a_x=a_y=:a)$.
Solid line is the exact result (Eq.~(\ref{eq:adiam})), 
dotted line is the IM (or DE) result (Eq.~(\ref{eq:adiam3})), 
and dashed line is the approximate result (Eq.~(\ref{eq:adiam4})).
}
\label{figradi}
\end{figure}

The \textcolor{black}{conditional}
probability density of the $k$th bead's position $\bm{R}_k$ 
when the strand's head and tail are pinned at $\bm{R}_0=\bm{0}$ and $\bm{R}=l\bm{\delta}_z$, respectively, 
is given by the product of two Green functions as
\begin{eqnarray}
p_k(\bm{R}_k|\bm{R}_0,\bm{R})
\propto G(\bm{R}_k,\bm{R}_0;k,0)G(\bm{R},\bm{R}_k;n,k).
\end{eqnarray}
The second moment of the $k$ th bead's position along the $x$-axis is calculated as
\begin{eqnarray}
\langle R_{k,x}^2\rangle &=&\frac{\int
R_{k,x}^2 ~p_k(\bm{R}_k|\bm{R}_0,\bm{R}) d\bm{R}_k}
{\int p_k(\bm{R}_k|\bm{R}_0,\bm{R}) d\bm{R}_k} \\
&=&\frac{b}{w_x\left(\coth\left[\frac{kbw_x}{3}\right]
+\coth\left[\frac{(n-k)bw_x}{3}\right]\right)}.
\label{eq:Rsecond}
\end{eqnarray}
The {\it radius of gyration} $a_x$ of the elliptical cross-section of the tube along the $x$ axis
can be estimated by taking the average of the last equation over all $k$, i.e.,
\begin{eqnarray}
a_x^2=\frac{1}{n}\int_0^{n} \langle R_{k,x}^2\rangle dk
= \frac{b}{2w_x}L\Bigl(\frac{n bw_x}{3}\Bigr)
\label{eq:adiam}
\end{eqnarray}
where $L(x):=\coth(x)-1/x$ is the Langevin function.
A similar relation holds for the $y$-component.
If inequality (\ref{eq:inequ}) is satisfied, then Eq.~(\ref{eq:adiam}) is written approximately as
\begin{eqnarray}
a_x^2 \simeq \frac{b}{2w_x}~~~~~~~(\mbox{for}~nbw_x\gg1).
\label{eq:adiam3}
\end{eqnarray}
This is the result that DE (for the isotropic case) \cite[Eq.~(A.7)]{Doi_Edwards_78b} 
and IM (for the anisotropic case) \cite[Eq.~(5)]{IM_JNNFM_1998} derived. 
Equation (\ref{eq:adiam3}) corresponds to the second moment for the 
Ornstein-Uhlenbeck process \cite{Uhlenbeck_PR_1930} in the steady state (when $n$ is interpreted as time).
In the absence of the confinement potential ($w_x=0$),
Eq.~(\ref{eq:adiam}) reduces to $a_x^2=l^2/18$ as it should.
By putting Eq.~(\ref{eq:monodenap}) into Eq.~(\ref{eq:adiam}), 
we obtain the approximate expression for the tube dimension as
\begin{eqnarray}
a_x^2\simeq \frac{b}{2w_x}
L\Bigl(\frac{lw_x}{\sqrt{2b\bar{w}+9b^2/l^2}}\Bigr).
\label{eq:adiam4}
\end{eqnarray}
See Fig.~\ref{figradi} where these approximate results are compared with the exact one.

\subsection{Monomer density}
\label{sebsec:monomerdensity}

\begin{figure}[t]
\begin{center}
\includegraphics*[scale=0.4]{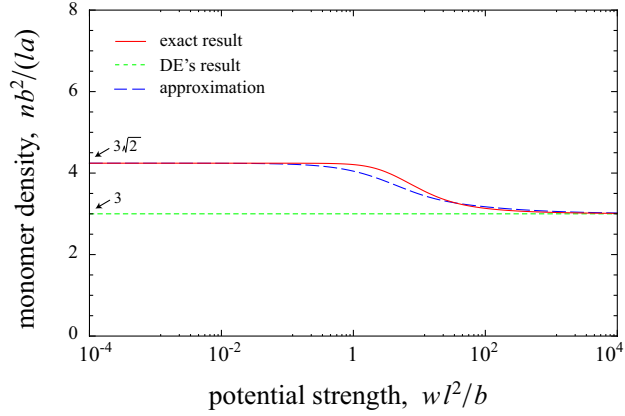}
\end{center}
\vspace*{-0.5cm}
\caption{
Monomer density (scaled by $a/b^2$) as a function of the strength of the confinement potential.
}
\label{figDE}
\end{figure}

DE showed that the monomer density per tube length, $n/l$, is of the order of $a/b^2$
by using the Green function (\ref{eq:GsolutionDE}) which is valid only when the confinement is strong~\cite{Doi_Edwards_78b}.
That is, by eliminating $w$ from two relations $n\sim 3l/(b\sqrt{2bw})$ (Eq.(\ref{eq:monoden1})) 
and $a\sim\sqrt{b/(2w)}$ (Eq.(\ref{eq:adiam3})) for the isotropic case, the monomer density can be estimated as $n/l\sim 3a/b^2$.
Interestingly, this prediction is valid even when the confinement is {\it weak}. 
By using two relations $n\simeq l^2/b^2$ (see Fig.~\ref{figmono}) and $a\simeq l/(3\sqrt{2})$ (see Fig.~\ref{figradi}) 
that hold in the weak confinement case, we obtain a relation $n/l\sim3\sqrt{2}a/b^2$.
The difference in these predictions is just a prefactor of order unity ($\sqrt{2}$), 
and we can therefore conclude that the DE prediction for the monomer density is appropriate 
for {\it any} strength of the confinement potential (see Fig. \ref{figDE}).

\subsection{Expression of the polarizability tensor}

Here we derive the polarizability tensor $\bm{A}$ from the tube segment.
In the calculation of the polarizability tensor, IM assumed 
that each link of the strand is not under the influence of the confinement potential 
(what we call the ``free confinement assumption").
On the other hand they {did not use} this free confinement assumption 
in the calculation of the stress tensor.
Here we avoid the free confinement assumption for the calculation of both tensors
because it seems inconsistent with Eq.~(\ref{eq:Green}) and is anyway unnecessary.

The origin of the polarizability and the stress of polymer melts is
the orientation distribution of the end-to-end vector of {\it units} of the chain.
Like IM, we assume that this unit is each link 
\textcolor{black}{(i.e., a segment of length $b$ along the continuum chain)}.
Validity of this assumption is discussed at the end of this section.
Then the local polarizability tensor from the $k$ th individual link is given by
\begin{eqnarray}
\bm{A}_{k}=c
\langle(\bm{R}_{k+1}-\bm{R}_{k})
(\bm{R}_{k+1}-\bm{R}_{k})\rangle
\label{eq:defpola}
\end{eqnarray}
where $c$ is a constant.
The brackets $\langle\cdots\rangle$ indicate taking the average for 
the conditional probability density 
that the $k$ th link's head (tail) is located at $\bm{R}_k$ ($\bm{R}_{k+1}$)
when the ends of the chain strand are fixed at {$\bm{R}_0=\bm{0}$ and $\bm{R}=l\bm{\delta}_z$} as before
(see Fig.~\ref{fig:tube})
\begin{eqnarray}
p_k(\bm{R}_k,\bm{R}_{k+1}|\bm{R}_0,\bm{R})
\propto G(\bm{R}_k,\bm{R}_0;k,0)G(\bm{R}_{k+1},\bm{R}_k;k+1,k)G(\bm{R},\bm{R}_{k+1};n,k+1).~~~~~~~
\end{eqnarray}
By using this conditional probability density, the $xx$-component of $\bm{A}_k$ is calculated as
\begin{eqnarray}
\langle(R_{k+1,x}-R_{k,x})^2\rangle
&=&\frac{\int
d\bm{R}_{k}\int
d\bm{R}_{k+1}~(R_{k+1,x}-R_{k,x})^2~p_k(\bm{R}_{k},\bm{R}_{k+1}|\bm{R}_0,\bm{R})}{\int
d\bm{R}_{k}\int
d\bm{R}_{k+1} ~p_k(\bm{R}_{k},\bm{R}_{k+1}|\bm{R}_0,\bm{R})} ~~~~~~~~\\
&=&\frac{b}{w_{x}}
\text{csch}\left[\frac{nbw_{x}}{3}\right]
\text{sinh}\left[\frac{bw_{x}}{3}\right]
\text{sinh}\left[\frac{k bw_{x}}{3}\right]
\text{sinh}\left[\frac{(k+1-n)bw_{x}}{3}\right] \nonumber\\
& &
\times\left(
\text{coth}\left[\frac{(k+1-n)bw_{x}}{3}\right]
-\text{coth}\left[\frac{k bw_{x}}{3}\right]
-2\text{tanh}\left[\frac{bw_{x}}{6}\right]
\right).
\label{eq:pol}
\end{eqnarray}
The $yy$-component satisfies a similar equation.

Now we calculate the principle components $\alpha_{xx},\alpha_{yy},\alpha_{zz}$ of 
the polarizability tensor $\bm{A}$ from the entire chain-strand confined 
in the tube segment\footnote{
In this paper we are following the IM assumption that
the stress and polarizability tensors originating from the chain strand 
in the tube segment have the principle directions,
one of which is along the central axis of the tube segment
($z$-axis) and the other two ($x$ and $y$ axes) are
along the principle directions of the ellipse cross section.}.
Since $n\gg1$, this is 
given by integrating $\bm{A}_k$ over $k$ for all links of the chain strand.
The $x$-component can be obtained from Eq.~(\ref{eq:pol}) as
\begin{eqnarray}
\alpha_{xx}=c\int_0^{n}  
\langle(R_{k+1,x}-R_{k,x})^2\rangle dk
=c\frac{nb^2}{3}\Pi_x
\label{eq:pora2kore0}
\end{eqnarray}
where
\begin{eqnarray}
\Pi_x:=\frac{6}{bw_x}
\sinh\left[\frac{bw_x}{6}\right]
\left(
\text{csch}\left[\frac{nbw_x}{3}\right]
\sinh\left[\frac{(2n-1)bw_x}{6}\right]
-\frac{3}{nbw_x}\cosh\left[\frac{bw_x}{3}\right]\sinh\left[\frac{bw_x}{6}\right]
\right).~~~~~~~~~~
\label{eq:Ppora2kore0}
\end{eqnarray}
The summation of $\bm{A}_k$ over $k$ yields a similar result but without the second term
in the parentheses of $\Pi_{x}$ \footnote{
\begin{eqnarray}
\alpha_{xx}&=&c\sum_{k=0}^{n-1} \langle(R_{k+1,x}-R_{k,x})^2\rangle
=c\frac{nb}{w_x}
\left(1-1/n
+\text{csch}\left[\frac{nbw_x}{3}\right]
\sinh\left[\frac{(n-1)bw_x}{3}\right]\right)
\tanh\left[\frac{bw_x}{6}\right]\\
&\simeq&
c\frac{2nb}{w_x}
\sinh\left[\frac{bw_x}{6}\right]
\text{csch}\left[\frac{nbw_x}{3}\right]
\sinh\left[\frac{(2n-1)bw_x}{6}\right]
\label{eq:pora2korefoot}
\end{eqnarray}
where $1/n$ in the parenthesis was neglected to obtain the final result.
}.
A comparison of this expression of $\alpha_{xx}$ with IM's result is discussed in App.~\ref{subsec:IM}.
The $y$-component has a similar expression.
On the other hand, the $z$-component is the polarizability from the Gaussian strand free from the confinement potential, i.e.,
\begin{eqnarray}
\alpha_{zz}
= c\frac{nb^2}{3}\Pi_z
\label{eq:appxpolz}
\end{eqnarray}
where
\begin{eqnarray}
\Pi_z:=1-\frac{1}{n}+\frac{3l^2}{n^2b^2}
\simeq 1+\frac{3l^2}{n^2b^2}.
\label{eq:Pappxpolz}
\end{eqnarray}
As a result, the polarizability tensor of a single tube segment 
aligned along the $z$-axis is
\begin{eqnarray}
\bm{A}= c\frac{nb^2}{3}\sum_{\beta=x,y,z}\Pi_\beta
\bm{\delta}_\beta\bm{\delta}_\beta.
\label{eq:Acorrect}
\end{eqnarray}
IM took an average of the polarizability tensor $\bm{A}$ for a single tube segment
over all tube-segment orientation and length,
and also multiplied it by the number density of the tube segment in the melt
to obtain the polarizability of the material. 
They did the same for the stress tensor.
In this paper, however, we don't perform these procedures for both quantities
because, for the purpose of confirming the SOR, it is 
enough to compare the polarizability and stress tensors
only for a single tube segment because the origin is the same.
Also IM replaced $n$ in the expression of the polarizability tensor with its
most probable value, which is, in our case, given as a solution of Eq.~(\ref{eq:monoden}).

Finally we comment briefly on the proper choice of the smallest unit 
responsible for the polarizability and the stress of the material.
{IM assumed that $b$ is the unit of polarizability, but here we consider the infinitesimal unit 
in accordance with the current, continuum picture of the chain.}
If we had discretized the chain by $\Delta$
as the unit of the polarizability, then the local polarizability from this portion would be given by
$\bm{A}_{k}^{(\Delta)}:=c\langle(\bm{R}_{(k+1)\Delta}-\bm{R}_{k\Delta})
(\bm{R}_{(k+1)\Delta}-\bm{R}_{k\Delta})\rangle$.
Integrating $\bm{A}_{k}^{(\Delta)}$ over all these portions $0\le k \le n/\Delta$,
the total polarizability tensor is obtained as
$\bm{A}^{(\Delta)}=c\frac{nb^2}{3}\sum_{\beta=x,y,z}\Pi_\beta^{(\Delta)}
\bm{\delta}_\beta\bm{\delta}_\beta$ where
\begin{eqnarray}
& &\Pi_{\beta(=x,y)}^{(\Delta)}=\frac{6}{bw_\beta\Delta }
\sinh\left[\frac{bw_\beta\Delta }{6}\right] \nonumber \\
& &~~~~~~~~~\times\left(
\text{csch}\left[\frac{nbw_\beta}{3}\right] 
\sinh\left[\frac{(2n-\Delta)bw_\beta}{6}\right]
-\frac{3}{nbw_x}\cosh\left[\frac{bw_x\Delta}{3}\right]\sinh\left[\frac{bw_x\Delta}{6}\right]\right),\\
& &\Pi_z^{(\Delta)}
=1-\frac{\Delta}{n}+\frac{3l^2}{n^2b^2}\Delta
\simeq 1+\frac{3l^2}{n^2b^2}\Delta.
\end{eqnarray}
The previous result (Eq.~(\ref{eq:Acorrect})) is recovered by putting $\Delta=1$.
On the other hand, in the limit of vanishing discretization $\Delta\to0$ for a fixed finite $b$ 
(or the continuum limit), the polarizability matrix becomes isotropic
\begin{eqnarray}
\lim_{\Delta\to0}\bm{A}^{(\Delta)}=c\frac{nb^2}{3}\bm{\delta},
\end{eqnarray}
where $\bm{\delta}$ is the unit tensor.
The isotropy comes from the elimination of the wavelength cutoff  
which makes the contour length of the strand infinite, as originally modeled.
Since we are not interested in this trivial result, 
we discretized the otherwise-continuous chain by putting $\Delta=1$ as did IM.

\subsection{Expression of the stress tensor}

Here we calculate the stress tensor,
and compare the result with the polarizability tensor derived above.

IM discussed that the contribution to the stress is two-fold;
one is the traction of the chain strand and the other is the confinement potential.
But they considered the traction only along the tube segment (i.e., $z$-axis),
which is consistent with their Green function.
Here we take the traction of all directions into account
to be consistent with our Green function.

According to the conventional theory for Gaussian chains \cite{Doi_TPD,Bird_DPL},
the stress tensor arising from the chain traction is given as the integral of the
second-moment tensor of the link's end-end vector over $k$.
The contribution from a single tube segment aligned along the $z$-axis is given by
\begin{eqnarray}
\bm{T}_c&=&
nk_{\rm B}T\sum_{\beta=x,y,z}\Pi_\beta\bm{\delta}_\beta\bm{\delta}_\beta
\label{eq:Tcorrectc}
\end{eqnarray}
where $\Pi_\beta$ is given by Eqs.~(\ref{eq:Ppora2kore0}) and (\ref{eq:Pappxpolz}).
Therefore $\bm{T}_c$ is proportional to $\bm{A}$ given by Eq.~(\ref{eq:Acorrect}),
indicating that the SOR holds if the stress does not include the contribution from the virtual springs 
representing the confinement potential (or `intrachain pressure').

On the other hand, the force arising from the virtual spring along the $\beta~(=x,y)$-axis is
$\bm{F}_\beta=-\frac{\partial V}{\partial \bm{R}_\beta}=-\frac{k_{\rm B}Tw_\beta^2}{3}\bm{R}_\beta$.
IM regarded this force as the origin of the pressure on the tube wall.
Thus the contribution to the stress tensor from the virtual springs is 
\begin{eqnarray}
\bm{T}_v&=&\sum_{\beta=x,y}
\int_0^{n}\!\!\langle F_{k,\beta}R_{k,\beta} \rangle
dk~\bm{\delta}_{\beta}\bm{\delta}_{\beta} \\
&=&-\sum_{\beta=x,y}
\frac{k_{\rm B}Tw_{\beta}^2}{3}
\int_0^{n}\!\!\langle R_{k,\beta}^2 \rangle 
dk~\bm{\delta}_{\beta}\bm{\delta}_{\beta}\\
&=&-k_{\rm B}T\sum_{\beta=x,y}
\frac{nbw_\beta}{6}
L\Bigl(\frac{nbw_{\beta}}{3}\Bigr)
\bm{\delta}_{\beta}\bm{\delta}_{\beta}
\label{eq:Tcorrectv}
\end{eqnarray}
where we used Eq.~(\ref{eq:adiam}).
Consequently, by adding these two parts, the total stress tensor for the tube segment is obtained as
\begin{eqnarray}
\bm{T}&=&\bm{T}_c+\bm{T}_v \nonumber \\
&=&nk_{\rm B}T
\left(
\sum_{\beta=x,y}
\left[ \Pi_\beta-\frac{bw_\beta}{6}L\Bigl(\frac{nbw_{\beta}}{3}\Bigr)\right]
\bm{\delta}_\beta\bm{\delta}_\beta
+\Pi_z \bm{\delta}_z\bm{\delta}_z
\right)
\label{eq:Tcorrecttotal2}
\end{eqnarray}

In the process of deriving Eq.~(\ref{eq:Tcorrecttotal2}), one {sees} that the stress tensor $\bm{T}$
is not proportional to the polarizability tensor $\bm{A}$ because of the presence of $\bm{T}_v$ in the stress. 
The virtual spring contributes directly to the stress, but not to the polarizability;
it affects the polarizability indirectly only through the Green function.
Thus the stress-optical rule is not generally obeyed in the present theoretical model of entangled polymer melts by IM.
It should be noted that the violation of SOR is not caused by the anisotropy of the potential 
but by the presence of the potential itself.
That is, the SOR is violated even if the confinement potential is isotropic $(w_x=w_y)$.
Also, one might expect from the IM prediction that the SOR holds if the confinement is very strong
because IM's prediction is based on the Green function that is appropriate in the strong confinement case.
However the SOR is violated even when the confinement is strong 
due to the presence of the confinement potential as shown in App.~\ref{app:compare}.

We make a final note about the interpretation of the wall pressure.
The sign of $\bm{T}_v$ was chosen to be the opposite of $\bm{T}_c$ 
so that $\bm{T}_v$ is an outward force (i.e., `pressure') toward the tube wall,
versus $\bm{T}_c$ representing the traction.
However it is unclear to us how to describe such a pressure by the confinement potential Eq.~(\ref{eq:potentiala})
which is, by definition, supposed to describe the attractive force toward the central axis of the tube.
We do not go into details of this issue, 
but rather just conclude this section by remarking that the SOR is not obeyed irrespective of the sign of $\bm{T}_v$.

\section{Conclusions}
\label{sec:conclusion}

We derived the Green function for an entangled polymer chain trapped 
in a tube having ellipsoidal cross-section described by an anisotropic harmonic potential.
Unlike the Green function derived by Doi-Edwards and that modified by Ianniruberto-Marrucci (IM),
ours satisfies the boundary condition along the axes perpendicular to the tube central axis.
The stress tensor and polarizability tensor derived from our Green function
without the free confinement assumption do not satisfy the stress-optical rule in the model proposed by IM.
The stress-optical rule is violated because the virtual springs of the confinement potential
contribute only to the mechanical stress tensor, not to the optical polarizability.
Thus the presence of the virtual spring itself, rather than the anisotropy of the spring potential,
is the source of the breakdown of the {stress-optical rule}.

\vspace*{1cm}

We are grateful to the Army Research Office (grants W911NF-08-2-0058, W911NF-09-1-
0378 and W911NF-11-2-0018) for financial support.

\appendix

\section{Comparison of IM's polarizability and {our} expression}
\label{subsec:IM}

With the help of the free confinement assumption, IM derived the polarizability 
from their Green function which is appropriate only for $nbw_x\gg1$ as \cite[Eq.~(17)]{IM_JNNFM_1998} 
\begin{eqnarray}
& &\alpha_{xx}^{(\rm{IM})}=c\frac{nb^2}{3}\frac{1}{1+bw_x/6}~~~~(\mbox{for}~1/n\ll bw_x).
\label{eq:poraIM}
\end{eqnarray}
On the other hand, {our} exact expression (Eq.~(\ref{eq:pora2kore0}) or Eq.~(\ref{eq:pora2korefoot}))
can be written approximately for the same condition as
\begin{eqnarray}
\alpha_{xx}\simeq c\frac{nb}{w_x}
\left(1-\text{e}^{-bw_x/3}\right)~~~~~~~~(\mbox{for}~1/n\ll bw_x).
\label{eq:pora2kore}
\end{eqnarray}
(It should be noted 
that Eq.~(\ref{eq:pora2kore}) divided by $n$
corresponds to the mean-square displacement of the particle
of the Ornstein-Uhlenbeck process in `steady state'
with the `lag time' corresponding to a single link \cite[p.77]{Gardiner}.)
{It is natural that Eq.(\ref{eq:poraIM}) and Eq.(\ref{eq:pora2kore}) are different
because these were derived under different assumptions.
But surprisingly, both equations are equal up to first order in $bw_x$
\begin{eqnarray}
\alpha_{xx}^{\rm (IM)}=\alpha_{xx}&\simeq&c\frac{nb^2}{3}
\left(1-\frac{bw_x}{6}\right)~~~~(\mbox{for}~1/n\ll bw_x \ll 1).
\label{eq:pora2ap}
\end{eqnarray}
This coincidence is due to the cancellation of errors in $\alpha_{xx}^{\rm (IM)}$ from 
(i) their Green function that gives rise to errors for small $bw_x$ and 
(ii) the free confinement assumption.
The higher-order terms of $\alpha_{xx}$ and $\alpha_{xx}^{\rm{(IM)}}$ do not agree
because of the free confinement assumption that makes the $w_x$-dependence weaker.
See Fig.~\ref{figax} where Eqs.~(\ref{eq:poraIM}), (\ref{eq:pora2kore}) and (\ref{eq:pora2ap}) are compared.
A similar discussion holds true for the $y$-component.

Indeed Eq.~(\ref{eq:pora2ap}) happens to hold in the condition $1/n\ll bw_x \ll 1$, 
but IM compared this result with the stress derived under the different conditions ($1/n\ll bw_x$ and $bw_x\gg1$)
in order to confirm the SOR. 
If these quantities are compared at the same conditions, we see that the SOR is violated as shown in App.\ref{app:compare}.

\begin{figure}[t]
\begin{center}
\includegraphics*[scale=0.4]{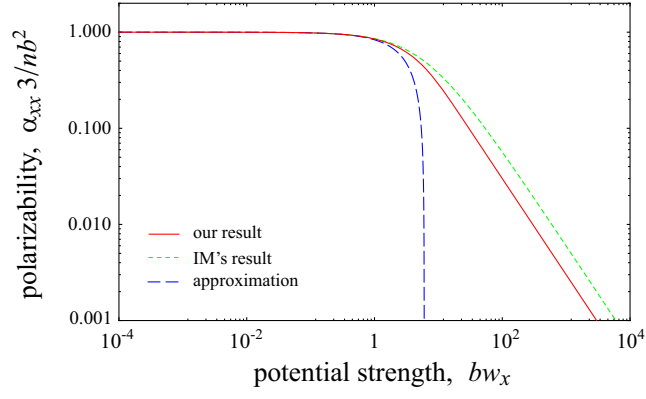}
\end{center}
\vspace*{-0.5cm}
\caption{
Polarizability plotted against the strength of the confinement potential.
Solid line is Eq.~(\ref{eq:pora2kore}),
dotted line is the result IM obtained,
and the dashed line is the approximate curve for both lines (Eq.~(\ref{eq:pora2ap})).
}
\label{figax}
\end{figure}

\section{Investigation of IM's discussion on the stress-optical rule}
\label{app:compare}

In this appendix, we consider in detail how we arrive at the conclusion that conflicts with IM.
For this purpose, {we focus on the `steady state' ($1\ll nbw_\beta$) as IM implicitly did.}
For clarity, we consider three cases where the confinement strength is 
(i) just lower bounded $1/n\ll bw_\beta<\infty$,
(ii) relatively weak $1/n\ll bw_\beta\ll1$, and 
(iii) very strong $1/n\ll bw_\beta$, $1\ll bw_\beta$.

(i) If $1/n\ll bw_\beta<\infty$,
Eq.~(\ref{eq:Tcorrecttotal2}) can be approximated as
\begin{eqnarray}
\bm{T}&\simeq n&k_{\rm B}T\left(\sum_{\beta=x,y}
\left[\frac{3}{bw_\beta}\left(1-\text{e}^{-bw_\beta/3}\right)
\underline{-\frac{bw_\beta}{6}}\right]
\bm{\delta}_\beta\bm{\delta}_\beta 
+\left(1+\frac{3l^2}{n^2b^2}\right)\bm{\delta}_z\bm{\delta}_z\right),
\label{eq:Tcorrecttotal3}
\end{eqnarray}
and also Eq.~(\ref{eq:Acorrect}) can be written as 
\begin{eqnarray}
\bm{A}\simeq c
\frac{nb^2}{3}
\left(
\sum_{\beta=x,y}
\frac{3}{bw_\beta}
\left(1-\text{e}^{-bw_\beta/3}\right)
\bm{\delta}_\beta\bm{\delta}_\beta
+\left(1+\frac{3l^2}{n^2b^2}\right)\bm{\delta}_z\bm{\delta}_z
\right)
\label{eq:Acorrect2}
\end{eqnarray}
where we used Eq.~(\ref{eq:pora2kore}).
SOR is violated due to the contribution to stress from the virtual springs (underlined term in Eq.~(\ref{eq:Tcorrecttotal3})).

(ii) If $1/n\ll bw_\beta\ll1$,
Eqs.~(\ref{eq:Tcorrecttotal3}) and (\ref{eq:Acorrect2})
can be expanded in a Taylor series in $bw_{\beta}$,
and consequently these can be decomposed into the isotopic and the anisotropic parts as
\begin{eqnarray}
\bm{T}\simeq nk_{\rm B}T\bm{\delta}+nk_{\rm B}T
\left(-\frac{bw_x}{3}\bm{\delta}_x\bm{\delta}_x
-\frac{bw_y}{3}\bm{\delta}_y\bm{\delta}_y
+\frac{3l^2}{n^2b^2}\bm{\delta}_z\bm{\delta}_z\right)
\label{eq:Tcorrecttotal4}
\end{eqnarray}
and
\begin{eqnarray}
\bm{A}&\simeq&c\frac{nb^2}{3}\bm{\delta}
+c\frac{nb^2}{3}
\left(-\frac{bw_x}{6}\bm{\delta}_x\bm{\delta}_x
-\frac{bw_y}{6}\bm{\delta}_y\bm{\delta}_y
+\frac{3l^2}{n^2b^2}\bm{\delta}_z\bm{\delta}_z\right),
\label{eq:Aappr}
\end{eqnarray}
respectively.
As discussed in App.~\ref{subsec:IM},
Eq.~(\ref{eq:Aappr}) for a single tube segment happens to correspond to the polarizability tensor 
that IM derived 
\cite[Eq.~(21)]{IM_JNNFM_1998}.
The anisotropic components of {Eqs.~(\ref{eq:Tcorrecttotal4}) and (\ref{eq:Aappr})}
are not proportional to each other, thereby violating the SOR.

(iii) If $1/n\ll bw_\beta$ and $1\ll bw_\beta$, then
the confinement is much stronger than the traction.
Therefore the contribution from the chain traction 
$\frac{3}{bw_\beta}(1-\exp[-bw_\beta/3])$ can be dropped
from Eq.~(\ref{eq:Tcorrecttotal3}),
and the first term $n$ in the parenthesis for the $z$-component can be neglected
compared with the $x$ and $y$ components $\frac{bw_\beta}{6}$ in Eq.~(\ref{eq:Tcorrecttotal3}).
Thus we have
\begin{eqnarray}
\bm{T}&\simeq&
nk_{\rm B}T
\left(
-\frac{bw_x}{6}\bm{\delta}_x\bm{\delta}_x 
-\frac{bw_y}{6}\bm{\delta}_y\bm{\delta}_y 
+\frac{3l^2}{n^2b^2}\bm{\delta}_z\bm{\delta}_z
\right)
\label{eq:Tcorrecttotal5}
\end{eqnarray}
This corresponds to the result that IM derived \cite[Eq.~(9)]{IM_JNNFM_1998}.
IM compared Eq.~(\ref{eq:Tcorrecttotal5}) to the anisotropic part of Eq.~(\ref{eq:Aappr})
to conclude that the SOR is obeyed even in the presence of the anisotropic confinement potential.
However, since the range of $w_\beta$ where each equation is approximately correct is not the same, 
one cannot verify SOR from such a comparison.
The SOR is not satisfied within the present condition because 
the polarizability tensor has only the $z$-component as
\begin{eqnarray}
\bm{A}\simeq c\frac{l^2}{n}\bm{\delta}_z\bm{\delta}_z.
\label{eq:Acorrect21}
\end{eqnarray}

Thus we conclude that the SOR is not obeyed in this model except the case 
where the potential is so weak as to be negligible compared with the chain traction,
and consequently both tensors are nearly isotropic: $\bm{T}\propto\bm{A}\propto\bm{\delta}$.
But we are not interested in this extreme case
because the chain strand is approximated by a free Gaussian chain without restriction
as shown in Fig.~\ref{figmono} and Fig.~\ref{figradi},
and therefore there is no surprise that the SOR holds.

\bibliographystyle{prsty}
\bibliography{jds,indeiSOR,Schieber_publications}

\end{document}